\documentclass[12pt,a4paper,final]{iopart}

\usepackage{iopams}  
\usepackage{graphicx}
\usepackage{multirow}
\usepackage{enumitem}
\usepackage{cleveref}
\usepackage[usenames, dvipsnames]{color}

\definecolor{mypink1}{rgb}{0.858, 0.188, 0.478}
\definecolor{mypink2}{RGB}{219, 48, 122}
\definecolor{mypink3}{cmyk}{0, 0.7808, 0.4429, 0.1412}
\definecolor{mygray}{gray}{0.6}
\usepackage[breaklinks=true,colorlinks=true,linkcolor=blue,urlcolor=blue,citecolor=blue]{hyperref}

\newcommand{\el}{El Ni\~{n}o}
\newcommand{\la}{La Ni\~{n}a}
\usepackage{setstack}
\usepackage{iopams}

\begin{document}

\title{Forecasting the magnitude and onset of \el ~based on climate network}

\author{Jun Meng$^{1,2}$, Jingfang Fan$^{1,2}$, Yosef Ashkenazy$^{1}$, Armin Bunde$^{3}$ and  Shlomo Havlin$^{2,4}$}
%\address{$^1$Department of Solar Energy \& Environmental Physics, Blaustein Institutes for Desert Research, Ben-Gurion University of the Negev, Midreshet Ben-Gurion 84990, Israel}
%\address{$^2$Department of Physics, Bar-Ilan University, Ramat-Gan 52900, Israel}
%%\ead{author.one@mail.com}

%\author[cor1]{Jingfang Fan$^{1,2}$}
\address{$^1$Department of Solar Energy \& Environmental Physics, Blaustein Institutes for Desert Research, Ben-Gurion University of the Negev, Midreshet Ben-Gurion 84990, Israel}
\address{$^2$Department of Physics, Bar-Ilan University, Ramat-Gan 52900, Israel}

%\author{Yosef Ashkenazy$^{1}$}
%\address{$^1$Department of Solar Energy \& Environmental Physics, Blaustein Institutes for Desert Research, Ben-Gurion University of the Negev, Midreshet Ben-Gurion 84990, Israel}

%\author{Armin Bunde$^{3}$}
\address{$^3$Institut f\"{u}r Theoretische Physik, Justus-Liebig-Universit\"{a}t Giessen, 35392 Giessen, Germany}
%\eads{\mailto{author.three@mail.com}, \mailto{author.three@gmail.com}}

%\author{Shlomo Havlin$^{2,4}$}
%\address{$^2$Department of Physics, Bar-Ilan University, Ramat-Gan 52900, Israel}
\address{$^4$Institute of Innovative Research, Tokyo Institute of Technology,
4259 Nagatsuta-cho, Midori-ku, Yokohama 226-8502, Japan}
\ead{j.fang.fan@gmail.com}

\begin{abstract}
\el ~is probably the most influential climate phenomenon on interannual time scales. It affects the global climate system and is associated with natural disasters; it has serious consequences in many aspects of human life. However, the forecasting of the onset and in particular the magnitude of \el ~are still not accurate enough, at least more than half a year ahead. Here, we introduce a new forecasting index based on climate network links representing the similarity of low frequency temporal temperature anomaly variations between different sites in the Ni{\~n}o 3.4 region. We find that significant upward trends in our index forecast the onset of \el ~approximately 1 year ahead, and the highest peak since the end of last \el ~in our index forecasts the magnitude of the following event. We study the forecasting capability of the proposed index on several datasets, \textcolor{mypink1}{including, ERA-Interim, NCEP Reanalysis I, PCMDI-AMIP 1.1.2 and ERSST.v5.}

\end{abstract}

%Uncomment for PACS numbers title message
% Keywords required only for MST, PB, PMB, PM, JOA, JOB? 
\vspace{2pc}

% Uncomment for Submitted to journal title message
\submitto{\NJP}
% Comment out if separate title page not required
%\maketitle

\section{Introduction}

\el ~Southern Oscillation (ENSO) is an inter-annual coupled ocean-atmosphere climate phenomenon~\cite{Dijkstra2005, Clarke2008, Cane2010}. \el ~is the warm phase of ENSO and is characterized by several degrees warming of the eastern equatorial Pacific ocean. It occurs every 3-5 years, and is regarded as the most significant climate phenomenon on decadal time scales. Among other factors, it affects the surface temperature, precipitation and mid-tropospheric atmospheric circulation over extended regions in America, Australia, Europe, India, and East Asia~\cite{Halpert1992,Diaz2001,Ropelewski1986,Kumar2006,Kushnir2007}. In particular, strong \el ~can trigger a cascade events that can affect many aspects of human life~\cite{Hsiang2011,Burke2015,Schleussner2016}. 

As a result of the environmental, economical, and social impacts of \el, intensive efforts have been undertaken to understand and eventually forecast \el ~\cite{McPhaden1998, Chen2008,Cane1986}. Extensive atmospheric and oceanic observations have been used to track variations in ENSO cycle, and \textcolor{mypink1}{many} complex computer models have been developed to forecast \el ~\cite{Barnett1988,Tang1997,Kirtman1997,Chen2004, Luo2008,Yeh2009,Galanti2003}. Still, reliable forecasts techniques for the onset and in particular the magnitude of \el ~with relatively long lead time (of more than half a year) are not fully satisfactory. We have just undergone one of the strongest \el~events since 1948, which started in the end of 2014 and ended in mid-2016~\cite{ENSO_report}. The onset of this event was predicted one year ahead using the network approach~\cite{Ludescher2014}.
\textcolor{mypink1}{Here, we develop a climate network based index to forecast the onset of \el ~approximately 1 year ahead (similar to \cite{Ludescher2014,Ludescher,Jun2017}). In particular our approach forecasts the magnitude of \el, once it begins.}

\section{Methodology}
The Oceanic Ni\~{n}o Index (ONI) is a standard index that is used to identify \el~\cite{oni}. It is the running 3-month mean sea surface temperature (SST) anomaly averaged over the Ni\~{n}o 3.4 region, based on 30 years periods, updated every 5 years. When the ONI exceeds $0.5^\circ$C for at least five consecutive months, the corresponding year is considered to be an \el ~year. We use the ONI (whose first value is at 1950) to estimate the accuracy of our predictions for \el ~events occurred after 1950.

We analyze the variability of the daily mean near surface (1000 hPa) air temperature fields of the ERA-Interim reanalysis~\cite{Dee2011}, the NCEP Reanalysis I~\cite{ncep}, the AMIP Sea Surface Temperature boundary condition data (current version: PCMDI-AMIP 1.1.3)~\cite{cmip6}, and the extended reconstructed Sea Surface Temperature v5 (ERSST.v5)~\cite{ersst} in the Ni{\~n}o 3.4 region (i.e., $5^\circ$S-$5^\circ$N, $120^\circ$W-$170^\circ$W) using a climate network approach~\cite{Yamasaki2008,Tsonis2008,Donges20091,Donges20092,Gozolchiani2011,
Wang2013,Zhou2015,Fan2017}. See Table \ref{comparison} (rows 1-5) for detailed information on the datasets. We find that the temporal variations of temperature anomaly (defined below in (i)) in different sites of the Ni{\~n}o 3.4 region become less coherent (more disordered) well before the onset of \el. In particular, the magnitude of the event is approximately proportional to the maximal degree of disorder (defined below in (ii)) that the Ni{\~n}o 3.4 region can reach before the onset of \el. We suggest a single index, the degree of disorder of the El Ni{\~n}o 3.4 region, that can forecast both the onset and magnitude of \el.

In the following, we first demonstrate the steps of the forecasting method we propose on 33 years (1984 to present) of the reanalysis data of the European Centre for Medium-Range Weather Forecasts Interim Reanalysis (ERA-Interim)~\cite{Dee2011}. We then examine the robustness and accuracy of the prediction method on longer periods using several other datasets (NCEP Reanalysis I~\cite{ncep}, PCMDI-AMIP 1.1.2~\cite{cmip6} and ERSST.v5~\cite{ersst}).

The daily mean near surface (1000 hPa) air temperature fields of the ERA-Interim reanalysis data have a spatial (zonal and meridional) resolution of $2.5^\circ \times 2.5^\circ$, resulting in 105 \textcolor{mypink1}{grid points} in the Ni\~{n}o 3.4 region. Different locations (grid points) in the Ni{\~n}o 3.4 region correspond to nodes in the local climate network, and the weight of links are determined by the similarities (defined below in (ii)) of the temporal temperature anomaly variations between pairs of nodes~\cite{Yamasaki2008,Fan2017}. The forecasting algorithm is as follows:
\begin{enumerate}[label=(\roman*)]
	\item\label{fi} At each node $k$ of the network, we calculate the daily atmospheric temperature
anomalies $T_k(t)$ (actual temperature value minus the climatological average which then is divided by the climatological standard deviation) for each calendar day. For the calculation of the climatological average and standard deviation, only past data up to the prediction date have been used. For simplicity leap days were excluded.  We have used the first 5 years of data (1979-1983) to calculate the first average value and start the prediction from 1984. 

	\item\label{fii} For obtaining the time evolution of the weight of the links between nodes $i$ and $j$ in the Ni{\~n}o 3.4 region, we follow~\cite{Ludescher,Jun2017,Yamasaki2008} and compute, for each month $t$ (the first day where the month starts) in the considered time span between Jan. 1, 1981 and Aug. 31, 2017, the time-delayed cross-correlation function defined as
\begin{equation}\label{c1}
C^{(t)}_{i,j}(-\tau)=\frac{\langle T_i^{(t)}(t) T_j^{(t)}(t-\tau) \rangle-\langle T_i^{(t)}(t)\rangle \langle T_j^{(t)}(t-\tau) \rangle }{\sqrt{\langle (T_i^{(t)}(t)-\langle T_i^{(t)}(t)\rangle)^2\rangle}\cdot\sqrt{\langle (T_j^{(t)}(t-\tau)-\langle T_j^{(t)}(t-\tau)\rangle)^2\rangle}},
\end{equation}
and
\begin{equation}\label{c2}
C^{(t)}_{i,j}(\tau)=\frac{\langle T_i^{(t)}(t-\tau) T_j^{(t)}(t) \rangle-\langle T_i^{(t)}(t-\tau)\rangle \langle T_j^{(t)}(t) \rangle }{\sqrt{\langle (T_i^{(t)}(t-\tau)-\langle T_i^{(t)}(t-\tau)\rangle)^2\rangle}\cdot\sqrt{\langle (T_j^{(t)}(t)-\langle T_j^{(t)}(t)\rangle)^2\rangle}},
\end{equation}
where the brackets denote an average over the past 365 days, according to
\begin{equation}
\langle f(t) \rangle=\frac{1}{365}\sum_{a=1}^{365}f(t-a).
\end{equation}
 We consider, for the daily datasets, time lags of $\tau \in [0, 200]$ days, where a reliable estimate of the background noise level can be guaranteed (the appropriate time lag is discussed in \cite{Guez2014}). For monthly updating datasets (PCMDI-AMIP 1.1.3 and ERSST.v5), the brackets denote an average over the past 12 months, according to $\langle f(t) \rangle=\frac{1}{12}\sum_{a=1}^{12}f(t-a).$ and we consider time lags of $\tau \in [0,6]$ months. The similarity between two nodes (the weight of the link) is determined by the value of the highest peak of the cross-correlation function, $C^{(t)}_{i,j}(\theta)$, where $\theta$ is the corresponding time lag at the peak. The degree of coherence/disorder of the Ni{\~n}o 3.4 region is quantified by the average value of all links at their peaks, i.e. 
\begin{equation}\label{sim}
C(t)=\frac{2}{N(N-1)}\sum_{i=1}^{N-1}\sum_{j=i+1}^{N}{C^{(t)}_{i,j}(\theta)},
\end{equation} 
where $N=105$ is the number of nodes in the Ni{\~n}o 3.4 region. Thus, higher values of $C(t)$ indicate higher coherence in the Ni{\~n}o 3.4 region.

We like to note that the strength of the link between nodes $i$ and $j$ is represented by the strength of the cross-correlation between the temperature records at the nodes, which is defined by~\cite{Ludescher,Gozolchiani2011} 
\begin{equation}
W_{i,j}^{(t)} = \frac{C_{i,j}^{(t)}(\theta) - \rm{\bf{E}}(C_{i,j}^{(t)})}{\sqrt{\rm{\bf{E}}(C_{i,j}^{(t)}-\rm{\bf{E}}(C_{i,j}^{(t)}))^2}},
\end{equation}
where $\rm{\bf{E}} (g)$ denotes the average over $401$ shifting days, according to
\begin{equation}
\rm{\bf{E}}(g)=\frac{1}{401}(\sum_{\tau=0}^{200}g(\tau)+\sum_{\tau=1}^{200}g(-\tau)).
\end{equation}
Thus, $W_{i,j}^{(t)}$ is high when the peak at $\tau=\theta$ is sharp and prominent, and it is low when the cross-correlation function $C_{i,j}^{(t)}(\tau)$ varies slowly with $\tau$. In ~\cite{Ludescher}, Ludescher, {\it et al.} introduced a 12-mo forecasting scheme based on the observation that the mean strength of links that connect the ``\el ~basin''(equatorial Pacific corridor) with the surrounding sites tends to increase about one year before the \el ~event. 

	\item\label{fiv} The forecasting index (FI) we propose here, is based on the temporal evolution of $ C(t)$ (defined in (ii) Eq. (\ref{sim})), representing the interactions or similarity (coherence) between the different sites within the  Ni{\~n}o 3.4 region. We define the FI as a function of months as follows, 
\begin{equation}\label{fif}
\rm{FI}(t)=\frac{1}{m+1}\sum_{a=0}^{m}\ln C(t-a)-\ln C(t) 
\end{equation}
where $m$ is the \textcolor{mypink1}{total} number of months before $t$ since Jan. 1981. We use a minus sign in the right hand side of Eq. \ref{fif} so that peaks in the FI will correspond to peaks in the ONI, see Fig.~\ref{predic-c-era}. We also use the $log$ in $C(t)$ instead of just $C(t)$ in order to make small variations of $C(t)$ to become more significant so that it will be seen more clearly in Fig.~\ref{predic-c-era}. We start to evaluate ${\rm FI}(t)$ from Jan. 1984. (For NCEP Reanalysis I, $m$ equals the number of months before $t$ since Jan. 1950, and the ${\rm FI}(t)$ starts from Jan. 1953; For PCMDI-AMIP 1.1.2, $m$ equals the number of months before $t$ since Jan. 1872, and ${\rm FI}(t)$ starts from Jan. 1950; For ERSST.v5, $m$ equals the number of months before $t$ since Jan. 1856, and  ${\rm FI}(t)$ starts from Jan. 1950) Thus, it follows that ${\rm FI}(t)$ increases ($C(t)$ decreases) when the Ni{\~n}o 3.4 region is less coherent or more disordered (due to the minus sign). ${\rm FI}(t)$ is calculated for each month (red dotted line in Fig.~\ref{predic-c-era}) and one can easily see that usually ${\rm FI}(t)$ increases well before the onset of \el, and decreases once \el ~begins. In other words, the temporal variations of temperature anomaly in different sites of the Ni{\~n}o 3.4 region become less coherent (more disorder) prior to \el, and start to synchronize once \el ~begins. In particular, we find that the more disordered the Ni{\~n}o 3.4 region is before \el, the higher is the magnitude of the approaching \el.
\end{enumerate}
\section{\textcolor{mypink1}{The forecasting algorithm using index FI}}

\begin{figure}[htb!]
\centering
\includegraphics[width=1.0\textwidth]{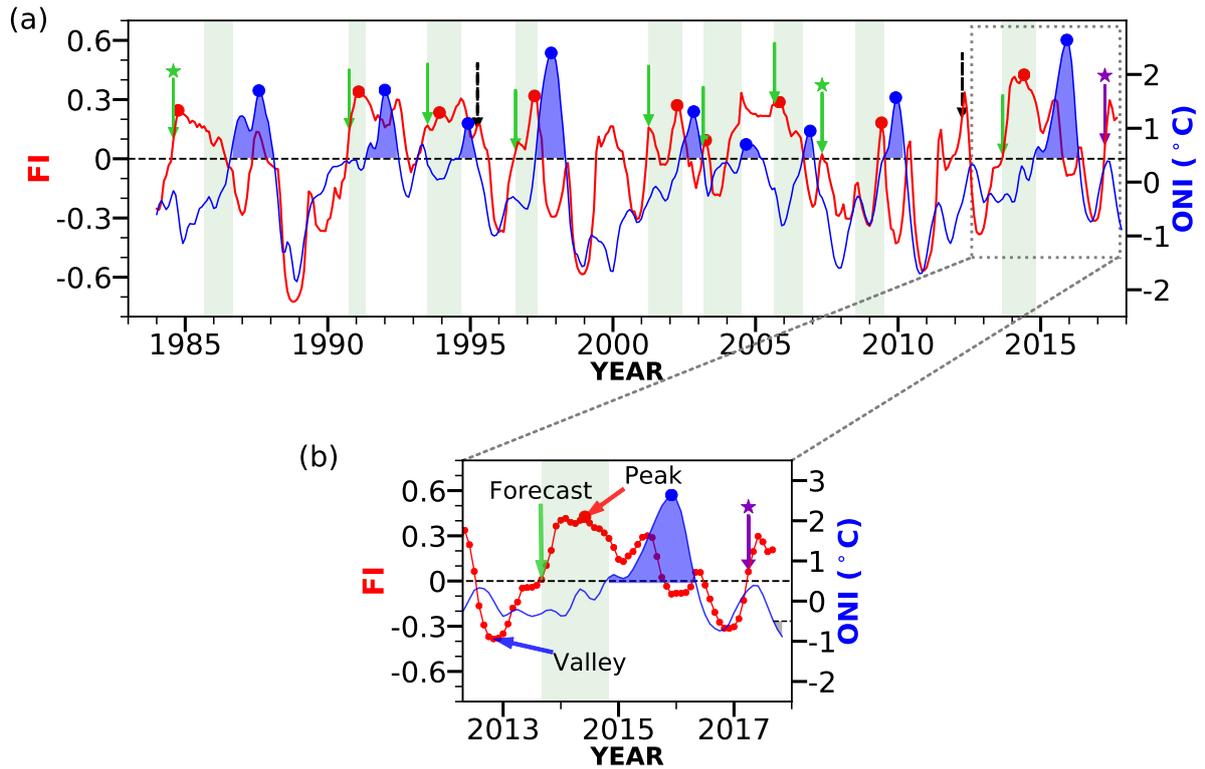}
%\centerline{\includegraphics[angle=0,width=1.0\columnwidth]{PREDICTION}}
\caption{\label{predic-c-era} The forecasting algorithm. (a) The Forecasting Index (${\rm FI}(t)$, red) and the ONI (blue). The blue shades under the ONI curve indicate the \el ~periods. True positive forecasts are marked by the green arrows, while false alarms are marked in black dashed arrows. The purple arrow is the newest potential alarm pending verification in the future. The green stars above some of the green arrows indicate the predictions after which \la ~occurred, leading to postponed \el ~prediction. The horizontal black dashed line indicates the ${\rm FI}=0$. The left edge of the light green rectangular shading indicates the time that ${\rm FI}(t)$ predicts \el ~and the right edge indicates the onset of \el, thus the width of the shading indicates the lead time before the onset of \el. The red dots indicate ${\rm FI}(t)$ values that are used to predict the magnitudes of \el ~and the blue dots indicate the actual magnitudes of \el. (b) A detailed view of ${\rm FI}(t)$ and the ONI since May, 2012. The green arrow (Forecast) indicates the true positive forecast of 2014-2016 \el ~event; the blue (Valley) and red (Peak) arrows indicate the nearest minimum before the prediction point and the peak in ${\rm FI}(t)$ is the value used to predict the magnitude of \el ~(blue dot). The purple star above the purple arrow indicates that we might be undergoing a new \la ~(the gray shades, the ONI has already been bellow $-0.5^\circ C$ for the last two months), and therefore an \el ~is forecasted to come within $18$ months after the end of the suspected ongoing \la. }
\end{figure}

Based on the above observation, we suggest the following algorithm to forecast simultaneously both the magnitude and onset of \el ~using ${\rm FI}(t)$. For demonstration see the example shown in Fig.~\ref{predic-c-era} (b). 
\begin{enumerate}
 	\item  To forecast the  magnitude, as soon as one month the ONI rises across $0.5^\circ C$ we regard the value of the highest peak of ${\rm FI}(t)$ (``Peak'', as indicated by the red points in Fig.~\ref{predic-c-era} (a) and the red arrow in (b)) since the end of last \el ~as an estimate (forecased magnitude) for \el ~strength (observed magnitude). However, if the peak value is negative or there is no peak during this period, we use zero as the forecasted magnitude and forecast a weak \el ~event (ONI$<1^\circ C$) (we counted the results of all the datasets we used, and find that the ratio of such events is $13\%$ on average of all the \el ~events, and most of them ($84\%$ on average of this kind of events) are indeed weak). In addition, we should clarify that if the ONI rises across but do not keep above $0.5^\circ C$ for at least five months, we do not have an \el ~event, thus the value of the highest peak is not a prediction of \el ~magnitude.
 	\item To forecast the onset, we track both ${\rm FI}(t)$ and the ONI, starting from the onset of the previous \el. If ${\rm FI}(t)$ increases from a local minimum (``Valley'', as indicated by the blue arrow in Fig.~\ref{predic-c-era} (b)) continuously for at least two months (time segment that yielded the best forecast), the time at which ${\rm FI}(t)$ exceeds 0 (if it is not during ongoing \el/\la ~period, i.e. $-0.5^\circ C<$ONI$<0.5^\circ C$) is considered as a potential signal for the onset of either \el ~or \la ~event within approximately the next 18 months (``Forecast'', as indicated by the green arrows in Fig.~\ref{predic-c-era}). Moreover, if \la ~is experienced within these 18 months, we forecast a new \el ~to occur within 18 months after the end of \la ~(the first month of ONI$>-0.5^\circ C$ after \la). Given the above, a true-positive prediction of \el ~is counted if within 18 months after the potential signal an \el ~occurs (``normal'', as indicated by the green arrows in Fig.~\ref{predic-c-era}), or a \la ~that followed by an \el ~in the next 18 months occurs (``delayed'', as indicated by the green arrows with stars on the top in Fig.~\ref{predic-c-era}); otherwise, a false alarm is counted. 
\end{enumerate}

\begin{figure}[htb!]
 \centering
\includegraphics[width=0.8\textwidth]{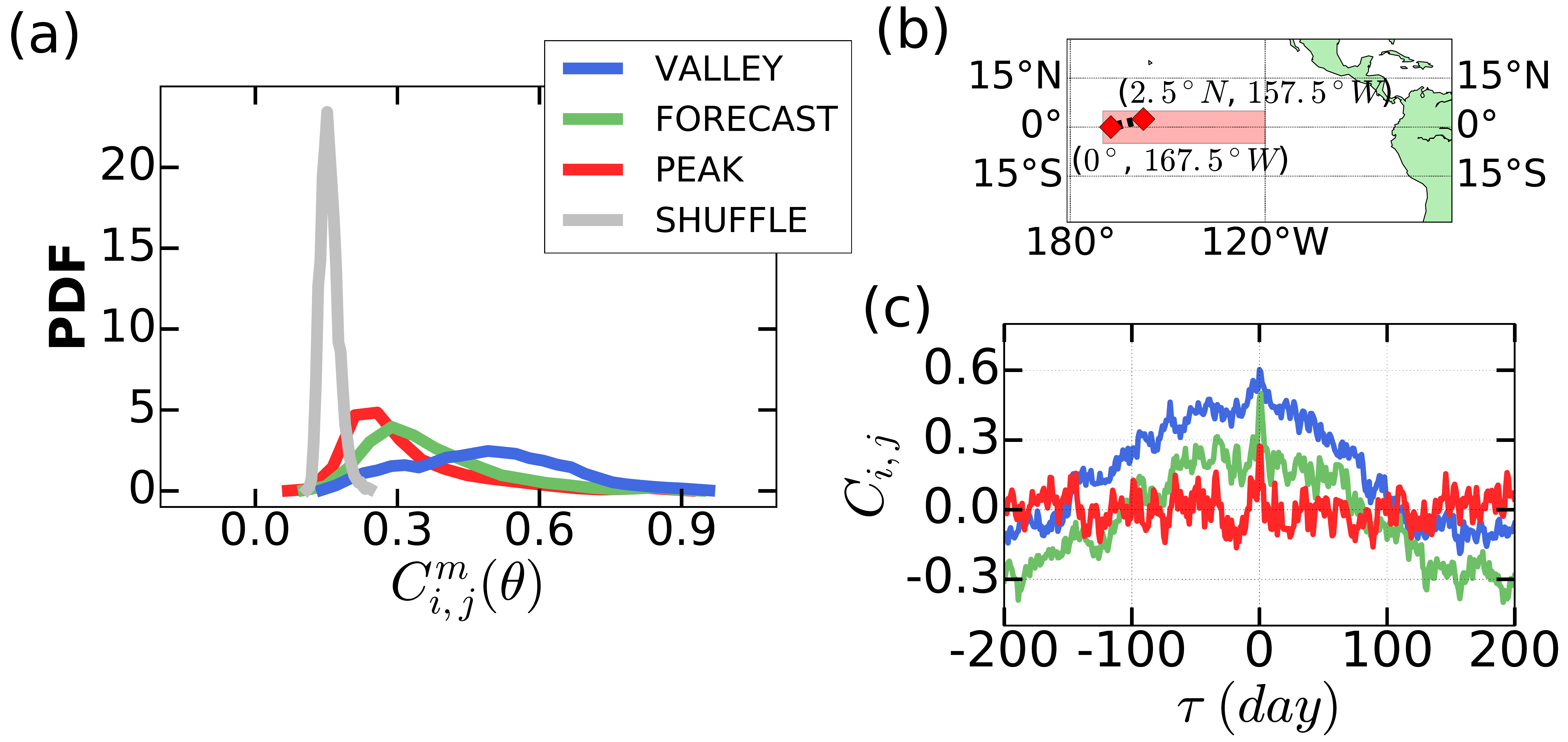}
%\centerline{\includegraphics[angle=0,width=0.85\columnwidth]{avesy}}
\caption{\label{pdf-cmax-era} (a) PDF of $C_{i,j}^{m}(\theta)$ for all links within Ni\~{n}o 3.4 region, in networks at VALLEY (blue), FORECAST (green), PEAK (red) months (see Fig. \ref{predic-c-era}), and for the random network (gray). (c) Typical  cross-correlation functions of a link (shown in (b)) within Ni\~{n}o 3.4 region (indicated by the light red shaded area) for months of VALLEY (blue), FORECAST (green), and PEAK (red) before the 2014-2016 \el ~event.} 
\end{figure}

Next, we elaborate on the reasoning behind our approach. In Fig.~\ref{pdf-cmax-era} (a), we plot the probability density function (PDF) of $C_{i,j}^{(t)}(\theta)$ for all links in network windows at which ``Valley'' ($m=Valley$, blue), ``Forecast'' ($m=Forecast$, green) and ``Peak'' ($m=Peak$, red) occur, respectively. We compare these PDFs with a PDF of random networks that are obtained by shuffling the order of the calendar days for each node within the Ni{\~n}o 3.4 region. We find the strongest correlations for the ``Valley'' periods (as the PDF is stretched toward higher values), then weaker correlations for the ``Forecast'' periods, and then the weakest correlations for the ``Peak'' periods (closest to the shuffled correlations). Thus, the Ni{\~n}o 3.4 network (region) becomes less coherent when progressing from ``Valley'' periods to the ``Peak'' periods. The order is reestablished towards the actual peak of \el. The evolution of the cross correlation of a typical link (shown in Fig.~\ref{pdf-cmax-era} (b)), before the onset of 2014-2016 \el ~event, is shown in Fig.~\ref{pdf-cmax-era} (c). The three cross correlation functions (blue, green, and red) correspond to the ``Valley'', ``Forecast'' and ``Peak'' points marked by blue, green and red arrows in Fig.~\ref{predic-c-era} (b). Consistently, we find that the maximal values of the cross correlation function, $C_{i,j}^{(t)}(\theta)$, decreases from time of ``Valley'' to ``Forecast'' time, then to the ``Peak'' time.

Moreover, while $C_{i,j}^{(t)}(\theta)$ is decreasing from Valley to Peak months, the strength of the link $W_{i,j}^{(t)}$~\cite{Ludescher} is increasing. This difference is probably due to the autocorrelation of the temperature anomaly variations in the Ni{\~n}o 3.4 region ~\cite{Guez2014}; see Fig. S1.

\section{Results}

\subsection{Forecasting the magnitude of \el}
\begin{figure}[htb!]
 \centering
\includegraphics[width=0.9\textwidth]{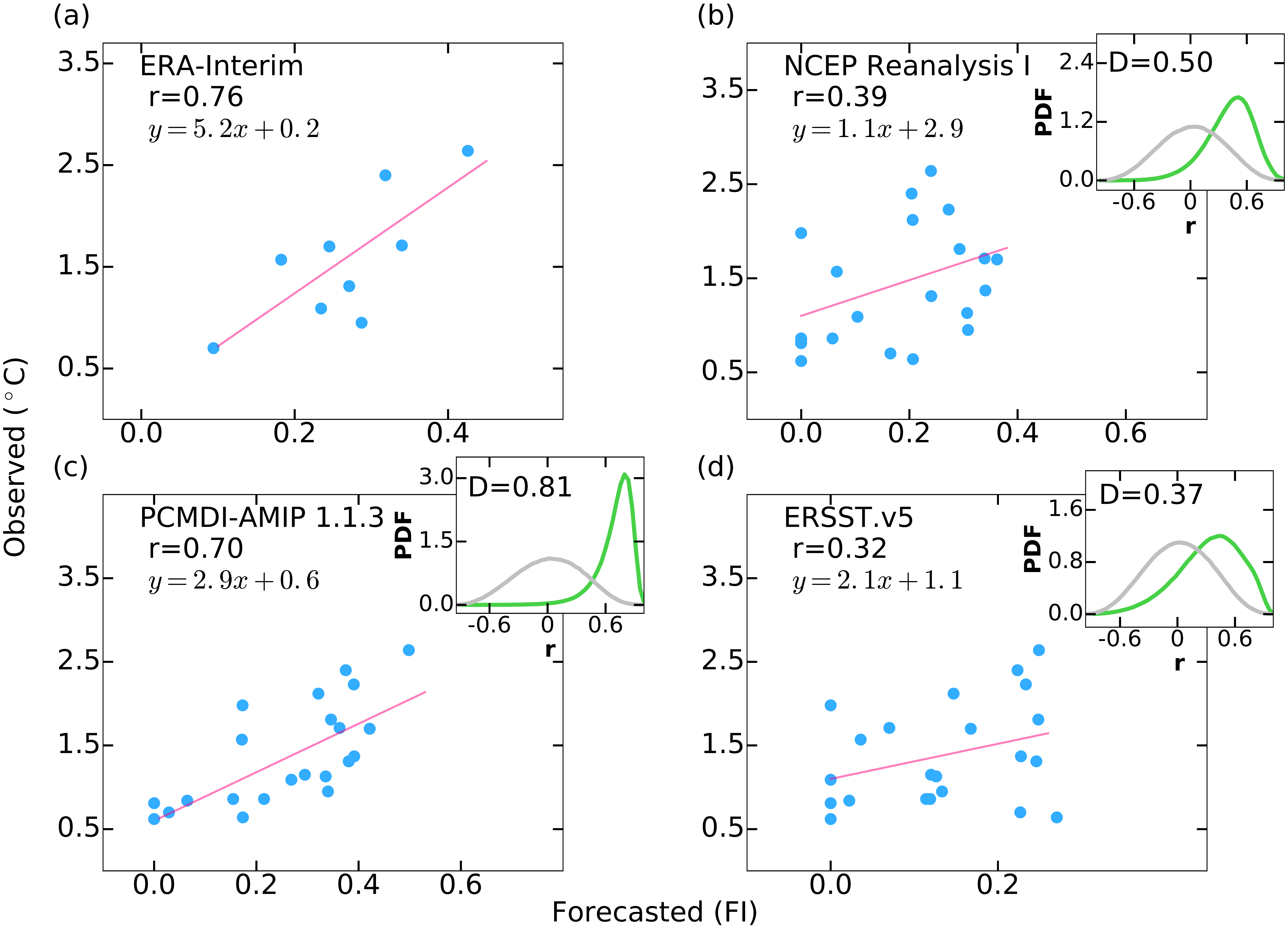}
%\centerline{\includegraphics[angle=0,width=1.0\columnwidth]{regress1950eraoni}}
\caption{ \label{strength} Scatter plots of the observed \el ~magnitude versus the forecast ones for different datasets (different panels) and the Kolmogorov-Smirnov statistic (inset of each panel). The different datasets are: (a) ERA-Interim~\cite{Dee2011}, (b) NCEP Reanalysis I~\cite{ncep}, (c) PCMDI-AMIP 1.1.1~\cite{cmip6}. and (d) ERSST.v5~\cite{ersst}. The red lines indicate the linear regression fits. We also show the corresponding correlation ($r$-values), Kolmogorov-Smirnov statistic ($D$), and the function of the fitting lines.}
\end{figure}

We now examine the accuracy and robustness of our forecast for the magnitude of \el ~events between 1950 and present (since the ONI begin from 1950), using several datasets. For this purpose, we plot the predicted magnitude versus the observed magnitude of \el ~(scatter plot), and use the Pearson correlation coefficient, $r$, to quantify the correlation. We present such scatter plots in Fig.~\ref{strength}.

Next, we apply the Kolmogorov-Smirnov test to quantify the significance of the relationship between the predicted and observed magnitude of \el; Fig.~\ref{strength} (insets). Each time we randomly choose ten events and calculate the correlation coefficient between their predicted and observed magnitudes; we repeated this procedure $1$ million times, and obtained the PDF of $r$-values for each dataset (colored by green in Fig.~\ref{strength}). For a comparison, we also consider random cases as follows. Each time we choose randomly ten predicted values and randomly ten observed values and then perform a linear regression between them; also here we have performed $1$ million selections, and obtain the PDF of $r$-values for each dataset (colored by gray in Fig.~\ref{strength}). Then we compare the PDFs of observed $r$-values to the random $r$-values using Kolmogorov-Smirnov statistic $D$~\cite{k-s}. For each dataset used here, $D$ is relatively large ($D\geq 0.37$), indicating significant difference between the observed and predicted \el ~magnitude. 
%However, for a few ($13\%$ on average) \el ~events for which the peak value is negative or there is no peak during this period, we find that such events have a high probability (of $83\%$ on average) to be weak events (peak value of the ONI $<1^\circ C$). Thus for these events, we assign their forecasted value to be zero. [THE LAST MODIFICATION IS A BIT PROBLEMATIC. ]

The results are summarized in Table \ref{comparison}, in the rows heading ``\el ~magnitude''. We note however, that the prediction of the magnitude of \el ~is performed at the actual onset of \el, which on average occurs about half a year prior to the peak of \el. 

\subsection{Forecasting the onset of \el}

Next we examine the forecasting power of the onset of \el. The results are summarized in Table \ref{comparison}, in the rows heading ``\el ~onset''. Here, the hit rate is defined as the hits (true positive prediction) divided by the number of \el ~events; the false alarm rate is defined as the number of false alarms divided by the number of years during which no \el ~started. The lead time equals the time from the potential signal (or the end of \la, if \la ~is experienced after the prediction) to the actual onset of \el ~(shaded areas in Fig.~\ref{predic-c-era}). 

Previous studies proposed various methods to forecast \el ~events. Some of these predict quite successfully the onset of \el, about one year in advance \cite{Ludescher, Jun2017}. We compare our prediction method to prediction of the 12-mo forecasting scheme based on climate-network approach~\cite{Ludescher} and to the prediction of state-of-the-art models---the COLA anomaly coupled model ~\cite{Kirtman2003} and the Chen-Cane model ~\cite{Chen2008}; for this purpose we use the operating characteristics (ROC)~\cite{roc}, see Fig. S3.
We use the Hit and False alarm rates as follows. The $Hit\ rate=\frac{hits}{hits+misses}$, where ``hits'' is the number of true positive prediction of \el. The $False\ alarm\ rate=\frac{false\ alarms}{false\ alarms+correct\ rejections}$, where the number of ``correct rejections'' equals the number of years where no \el ~started and no false alarm appeared in the past $18$ months before the year. The resulting hit rate of our approach is $\geq 0.81$ and the false alarm ratio is $\leq 0.24$. For prediction lead time of 12 months the hit rate is $<0.4$ for the COLA model~\cite{Kirtman2003} and $<0.45$ for the Chen-Cane model~\cite{Chen2008} with false alarm ratio of $\sim 0.2$. The hit rate for the network approach in~\cite{Ludescher} is $0.667$ and false alarm rate is $0.095$. 

The prediction scheme we proposed here improves the prediction of the onset of \el. An additional and also the most important advantage of the prediction scheme we propose is that it provide prediction both for the magnitude and the onset of \el ~ based only on the temperature variability and their coherence in the Ni{\~n}o 3.4 region.

\begin{table}[htbp]
\small
\setlength\tabcolsep{2pt} 
\resizebox{1.0\textwidth}{!}
%\begin{table}
%\scalebox{1}
{
\begin{tabular}{c|c||c|c|c|c}
\hline
\multicolumn{2}{c||}{DATA}&ERA-Interim~\cite{Dee2011}&NCEP Reanalysis I~\cite{ncep}&PCMDI-AMIP 1.1.3~\cite{cmip6}&ERSST. v5~\cite{ersst}\\
\hline
\multicolumn{2}{c||}{Type of data}&\multicolumn{2}{c|}{Daily near surface (1000 hPa) air temperature}&\multicolumn{2}{c}{Monthly sea surface temperature}\\
\hline
\multicolumn{2}{c||}{The first year of data}&1979&1948&1870&1854\\
\hline
\multicolumn{2}{c||}{The first year of the {\rm FI}}&1984&1953&1950&1950\\
\hline
\multicolumn{2}{c||}{Resolution of the data}&$2.5^\circ \times 2.5^\circ$&$2.5^\circ \times 2.5^\circ$&$1^\circ \times 1^\circ$&$2^\circ \times 2^\circ$\\
\hline\hline
\multirow{2}{*}{\el ~magnitude}
&r-value&0.76&0.39&0.70&0.32\\
\cline{2-6}
&$D$&*&0.50&0.81&0.37\\
\cline{2-6}
\hline
\multirow{3}{*}{\el ~onset}&Hit rate&$(7+2)/9=1$&$(13+4)/21\approx 0.81$&$(18+3)/22\approx 0.95$&$(19+2)/22\approx 0.95$\\
\cline{2-6}
&False alarm rate&$2/23\approx 0.09$&$8/43\approx 0.19$&$9/42\approx 0.21$&$10/41\approx 0.24$\\
\cline{2-6}
&Lead time (month)&$12.2 \pm 2.6$&$9.3 \pm 4.9$&$9.1 \pm 4.7$&$6.9 \pm 4.1$\\
\hline
\end{tabular}
}
\caption{\label{comparison} Summary of the information (rows 1-5) and the forecasting power (rows 6-10) for different datasets.  For the first two datasets, we start the prediction from the sixth year from the beginning of the data. For the last two datasets, we forecasts all the \el ~events since 1950, since there is enough historical data to calculate the first value of ${\rm FI}(t)$ in Jan. 1950. In all the datasets, ${\rm FI}(t)$ is only based on the past data, from the beginning of the data. The ``Resolution of the data'' refers to the spatial (zonal and meridional) resolution of the data. In the row of ``$D$'' , the symbol ``*'' indicates that there are not enough events in the datasets to perform the Kolmogorov-Smirnov statistic. \textcolor{mypink1}{The numerator of hit rate is composed of two parts, ``normal''$ +$``delayed'',  the denominator is the number of all the \el ~events for each dataset. The numerator of the false alarm rate is the number of false alarms,  the denominator is the sum of number of false alarms and the ``correct rejections'' which equals the number of years where no \el ~started and no false alarm appeared in the past $18$ months before the year.} In the columns of lead time, we show the mean value $\pm$ 1 standard deviation in units of months.  For more details on the above results see Figs. S2 in the SI.}
\end{table}

\section{Summary} 
In summary, we introduce a new forecasting index (FI) that is based on climate networks which accurately and  simultaneously forecasts both the onset and magnitude of \el. The performance of the FI is examined successfully on several datasets. Our forecasting algorithm is based on the finding that the similarity or the coherence of low frequency temporal variability of temperature anomaly between different sites (strength of links) in the Ni{\~n}o 3.4 region decreases well before \el ~and increases at the onset of \el. The magnitude of the predicted \el ~is positively related with the highest peak in the FI during the period between the end of last \el ~and the onset of the new one. The results presented here indicate an important characteristic of the phase of the ENSO cycle, i.e., significant increase of disorder occurs in the Ni{\~n}o 3.4 region well before the onset of \el. The relationship between \el ~and the variation of the degree of disorder in the Ni{\~n}o 3.4 region may be further explained by defining an entropy based on the coherence of temperature variations in different sites of the Ni{\~n}o 3.4 region, which oscillates periodically with the ENSO cycle. There is surely a room of further improvement of the forecasting algorithm proposed here, probably with combination with other forecasting techniques and models. 

\section{Acknowledgments}

We thank Kai Xu for helpful discussions and suggestions.
We acknowledge the Israel-Italian collaborative
project NECST, the Israel Science Foundation, ONR, Japan Science Foundation, BSF-NSF, and
DTRA (Grant No. HDTRA-1-10-1-0014) for financial support. J.F thanks the fellowship program
funded by the Planning and Budgeting Committee of the Council for Higher Education of Israel.

\end{document}